\documentclass[12pt]{article}
\usepackage{amsmath}
\begin{document}
\title{Extending additivity from symmetric to \\ asymmetric channels}
\author{Motohisa Fukuda \\
Statistical Laboratory, \\
Centre for Mathematical Sciences, \\
University of Cambridge, \\
Wilberforce Road, \\
Cambridge, CB3 0WB, UK \\
m.fukuda@statslab.cam.ac.uk}
\maketitle

\begin{abstract}
We prove a lemma which allows one to extend results about
the additivity of the minimal output entropy from
highly symmetric channels to a much larger class.
A similar result holds for the maximal output $p$-norm.
Examples are given showing its use in a variety of situations.
In particular, we prove the additivity and the multiplicativity
for the shifted depolarising channel.
\end{abstract}

A natural and important class of measures of the noisiness of a
quantum channel are based on optimal output purity, i.e., 
how close can the output be to a pure state as measured by 
the minimal ouput entropy (MOE)
or the maximal output $p$-norm.    
It is then natural to ask if
a tensor product of two channels can ever be less noisy in the sense
that some entangled input can have its output closer to a pure state
than the product of the optimal inputs of the two channels.    
This leads to the conjectures of the additivity of the MOE
and the multiplicativity of the maximal output $p$-norm. 
Although the additivity of the MOE 
has been shown [1] to be
equivalent globally to several other fundamental
conjectures in quantum information theory;
the additivity of the Holevo capacity,
the additivity of the entanglement of formation and
the strong superadditivity of the entanglement of formation,
we consider only the output purity.       

Intuitively, one would expect that entanglement would be more likely
to enhance output purity for highly symmetric channels
than for non-unital or asymmetric ones. 
Symmetric channels allow one to
construct maximally entangled states
as superpositions of products of optimal inputs,
but this can not be done for a channel with a unique optimal input.   
Curiously, however, most channels for which 
the additivity or
the multiplicativity has been proven are highly symmetric:
unital qubit channels [2],
the depolarising channel [3],[4], 
the Werner-Holevo channel [5],[6],[7],
and the transpose depolarising channel [8],[9].
These proofs exploit
the symmetric structures of these channels.
By contrast,
asymmetric channels have been extremely resistant
to proofs.
Indeed, apart from entanglement-breaking channels [10] and
one modification of the Werner-Holevo channel [11], 
the additivity has not
been proven for any non-unital channels 
and the multiplicativity   
only at $p=2$ in a few additional cases [12],[13].

In this note we present a lemma which allows one to prove 
the additivity of the MOE and 
the multiplicativity of the maximal output $p$-norm 
for a large class of channels. 
These include the shifted depolarising channel 
introduced in [12],
as well as many other types of 
non-unital channels for which our methods can
show that  multiplicativity holds for values of $p$ 
which include the interval $[1,2]\subseteq [1,\infty]$.
In essence, the lemma allows the extension of 
the additivity and the multiplicativity from
highly symmetric channels to asymmetric ones.  

Let us give some basic definitions.
A quantum state is represented as a positive 
semidefinite operator $\rho$ of trace one in a Hilbert 
space $\mathcal{H}$ of dimension $d$; this is called a 
density operator. The set of density operators in $\mathcal{H}$ 
is written as ${\mathcal D}(\mathcal{H})$.
A channel $\Phi$ on $\mathcal{H}$ is 
a completely positive  
trace-preserving (CPTP) map acting on $\mathcal{D}(\mathcal{H})$.
A channel $\Phi$ is called unitarily covariant if
for any unitary operator $U$ there exists a unitary operator $V$ such that
\begin{align}
\Phi(U \rho U^\ast)=V\Phi(\rho)V^\ast
\end{align}
for $\forall \rho \in {\mathcal D}(\mathcal{H})$.
The MOE of $\Phi$ is defined as
\begin{align}
S_{\min}(\Phi):=\inf_{\rho \in {\mathcal D}(\mathcal{H}) }S(\Phi(\rho)),
\end{align}
where $S$ is the von Neumann entropy:
$S(\rho)=-{\rm{tr}}[\rho\log\rho]$.
The maximal output $p$-norm of $\Phi $ is defined as
\begin{align}
\nu_p(\Phi):=\sup_{\rho \in {\mathcal D}(\mathcal{H})}\|\Phi(\rho)\|_p,
\end{align}
where $\|\;\;\|_p$ is the Schatten $p$-norm: $\|\rho\|_p=({\rm{tr}}|
\rho|^p)^{\frac{1}{p}}$.

The additivity conjecture of the MOE is that
for channels $\Phi$ and $\Omega$
\begin{align} \label{add}
S_{\min}(\Phi \otimes \Omega)= S_{\min}(\Phi)+S_{\min}(\Omega).
\end{align}
Note that the bound $S_{\min}(\Phi \otimes \Omega) \leq  S_{\min}(\Phi)+S_{\min}(\Omega)$ 
is straightforward.
The multiplicativity conjecture is that
for channels $\Phi$ and $\Omega$
\begin{align} \label{mult}  
\nu_p(\Phi \otimes \Omega)= \nu_p(\Phi)\nu_p(\Omega),
\end{align}
for any $p \in [1,2]$.
Note that the bound 
$\nu_p(\Phi \otimes \Omega) \geq  \nu_p(\Phi)\nu_p(\Omega)$ 
is straightforward.
Note also that 
if there exists a sequence $\{p_n\}$ with $p_n \searrow 1$ for which
the multiplicativity (\ref{mult}) holds then
the additivity of the MOE (\ref{add}) also holds [14].

{\bf Lemma.} {\sl
Suppose we have two channels on $\mathcal{H}$:
a unitarily covariant channel $\Psi$ and
a channel $M$ such that $M(\rho_0)$ is of rank one
for some state $\rho_0 \in {\mathcal D}(\mathcal{H})$.
Take another Hilbert space $\mathcal{K}$
and a channel $\Omega$ on $\mathcal{K}$.\\
1) If the maximal output $p$-norm of $\Psi \otimes \Omega$ is multiplicative
then so is that of $(\Psi \circ M)\otimes\Omega$ 
(for the same $p$).\\
2) If the MOE of $\Psi \otimes \Omega$ is additive 
then so is that of $(\Psi \circ M)\otimes\Omega$. }

{\bf Proof.}
We first claim that the maximal output $p$-norm of $\Psi \circ M$ satisfies
\begin{align} \label{mop} 
\nu_p(\Psi \circ M)
&=\sup_{\rho \in {\mathcal D}(\mathcal{H})}\| \Psi(M(\rho)) \|_p \notag\\
&=\sup_{\sigma \in M({\mathcal D}(\mathcal{H}))}\|\Psi(\sigma)\|_p 
=\nu_p(\Psi).
\end{align} 
Since $\Psi$ is unitarily covariant
the maximal output $p$-norm of $\Psi$ is attained
at any state of rank one.
On the other hand,
$M({\mathcal D}(\mathcal{H}))$, the image of ${\mathcal D}(\mathcal{H})$ by $M$, has 
such a state $M(\rho_0)$ for some $\rho_0 \in {\mathcal D}(\mathcal{H})$.
This verifies (\ref{mop}).
Next, take any state $\hat{\rho}\in {\mathcal D}(\mathcal{H} \otimes \mathcal{K})$ then
\begin{align}
\|( ( \Psi \circ M)\otimes \Omega ) (\hat{\rho})\|_p
&=\|(\Psi \otimes \Omega)((M \otimes {\bf 1}_{\mathcal{K}})(\hat{\rho}))\|_p  \notag\\
&\leq \nu_p(\Psi)\nu_p(\Omega)
=\nu_p(\Psi\circ M)\nu_p(\Omega).
\end{align}
Therefore we have
\begin{align}
\nu_p((\Psi \circ M)\otimes \Omega )
\leq \nu_p(\Psi\circ M)\nu_p(\Omega).
\end{align}
Similarly, 
for any state $\hat{\rho}\in {\mathcal D}(\mathcal{H} \otimes \mathcal{K})$ 
\begin{align}
S(((\Psi \circ M)\otimes \Omega)(\hat{\rho}))
&\geq S_{\min}(\Psi)+S_{\min}(\Omega)\notag\\
&=S_{\min}(\Psi \circ M )+S_{\min}(\Omega).
\end{align}
Hence we have
\begin{align}
S_{\min}((\Psi \circ M)\otimes \Omega)
\geq S_{\min}(\Psi \circ M )+S_{\min}(\Omega).
\end{align}
Combining these inequalities with those noted above in the
reverse direction completes the proof.  ~~~~~ QED

{\bf Remark.}
The conditions on $\Psi$ and $M$ can be weakened to
verify that
a maximiser of $ \|\Psi(\rho)\|_p$ 
(or a minimiser of $S(\Psi(\rho))$)
coincides with an output of $M$.

We now use the lemma to obtain some new results.
Our first group of examples are based on 
the depolarising channel defined by
\begin{align}
\Delta_\lambda^{(d)}(\rho)=\lambda\rho+(1-\lambda)\frac{1}{d}I.
\end{align}
Here $\lambda\in [-1/(d^2-1),1]$, 
$d$ is the dimension of the signal Hilbert space ${\mathcal H}$ and 
$I$ is the identity operator. 
The additivity of the MOE (\ref{add}) and 
the multiplicativity (\ref{mult}) with $p \in [1,\infty]$
have been proven for
$\Delta_\lambda^{(d)} \otimes \Omega$
with $\Omega$ arbitrary [3].
When $M$ is a channel with an output of rank one, 
consider the class of channels:
\begin{align} \label{ch0}
\Phi(\rho)=(\Delta_\lambda^{(d)} \circ M)(\rho)
= \lambda M(\rho)+(1-\lambda)\frac{1}{d}I.
\end{align}
Since the depolarising channel is unitarily covariant 
the additivity of the MOE (\ref{add}) and 
the multiplicativity (\ref{mult})
for $p \in [1,\infty]$
hold for the product of 
a channel of the form (\ref{ch0}) and 
an arbitrary channel.
We discuss examples of this class below.
\\
{\bf Example a).}
The shifted depolarising channel is defined by
\begin{align}
\Phi(\rho)=a\rho+b|\phi\rangle\langle\phi|+c\frac{1}{d}I.
\end{align}
Here $|\phi\rangle\langle\phi|$ is 
a fixed state of rank one,
$a,b,c \geq  0$ and $a+b+c=1$.
The multiplicativity (\ref{mult})
for $p=2$ was proven 
for the shifted depolarising channel [12],[13].
To see that this channel has the form (\ref{ch0}),
define a channel $M$ by
\begin{align} \label{shift}
M(\rho)=\frac{a}{a+b}\rho
+\frac{b}{a+b}|\phi\rangle\langle\phi|,
\end{align}
and put $\lambda=a+b$.
\\
{\bf Example b).}
Let $M$ be a channel of the form:
\begin{align}
M(\rho)= \frac{1}{\lambda}\sum_k \lambda_k V_k \rho V_k^\ast 
\end{align}
Here $\lambda_k >0$, $\lambda = \sum_k \lambda_k \leq 1$, and
$V_k$ are unitary operators having a common eigenvector.
Then define a channel
\begin{align}
\Phi(\rho) = \sum_k \lambda_k V_k \rho V_k^\ast + (1-\lambda)\frac{1}{d}I.
\end{align}
This channel was introduced in [15],
where both the additivity of the MOE (\ref{add}) and 
the multiplicativity (\ref{mult})
were proven by a different method.

Next we consider qubit channels.
The additivity of the MOE (\ref{add}) and 
the multiplicativity (\ref{mult})
were proven for unital qubit channels [2].
We use our results to extend this 
to some non-unital qubit channels.
Recall that any qubit state can be written as 
\begin{align} \label{bloch}
\rho=\frac{1}{2}\left[I + \sum_{k=1}^3 w_k \sigma_k \right].
\end{align}
Here ${\bf w} \in {\bf R^3}$ with $|{\bf w}|\leq 1$
and $\sigma_k$ are pauli matrices.
Note that $\rho$ is of rank one if and only if
$|{\bf w}| = 1$.
Let $ \Upsilon_{{\bf x},t} $ denote the channel:
\begin{align} \label{ch1}
\Upsilon_{{\bf x},t}(\rho)=\frac{1}{2}\left[ I +
x_1 w_1 \sigma_1 + x_2 w_2 \sigma_2
+ (t + x_3 w_3) \sigma_3 \right].
\end{align}
Here we assume for simplicity
$x_k > 0$ for $k=1,2,3$ and $t>0$.
Then we need the conditions:
$x_1,x_2,x_3+t <1$, and
\begin{align} \label{nasc1}
(x_1 \pm  x_2)^2 \leq (1\pm x_3)^2-t^2.
\end{align}
These condition are necessary and sufficient
for a map of the form (\ref{ch1}) to be CPTP [16],[17].
The qubit depolarising channel $ \Delta_\lambda^{(2)} $ 
belongs to this class
with $x_1=x_2=x_3=\lambda$ and $t=0$.
In the Bloch sphere representation (\ref{bloch})
the image of the sphere by a channel of the form (\ref{ch1}) is
an ellipsoid with axes of length
$x_k$ and centre $(0,0,t)$.
\\
{\bf Example c).}
Let $M =\Upsilon_{{\bf x},t}$ with
\begin{align} \label{extc}
x_3=x_1x_2,\;\; t^2 = (1-x_1^2)(1-x_2^2)\neq 0
\end{align}
then
$M$ is an extreme point of the set of qubit CPTP maps and
has an output of rank one.
In fact, it has two such outputs [16] unless $x_1 = \pm x_2$,
when it reduces to the amplitude damping channel.
In both cases the lemma can be applied to the channel:
\begin{align}
\qquad \Upsilon _{{\bf y},u}(\rho)
=(\Delta_\lambda^{(2)}\circ M)(\rho)
=\frac{1}{2}\left[I+ \lambda x_1 w_1 \sigma_1
+ \lambda x_2 w_2 \sigma_2 + \lambda(t + x_3w_3)\right].
\end{align}
to show that the additivity of the MOE (\ref{add}) and 
the multiplicativity (\ref{mult}) for $p \in [1,\infty]$ hold
for $\Upsilon _{{\bf y},u} \otimes \Omega$
with $\Omega$ arbitrary.
Here the conditions (\ref{extc}) imply
$y_1 y_2 =\lambda y_3 $, hence $|y_1 y_2| \leq |y_3|$, and 
$u^2=(y_2^2-y_3^2)(y_1^2-y_3^2)/y_3^2$.
In this example the shift $(0,0,u)$ is
in the direction of the shortest axes of the ellipsoid.
\\
{\bf Example d).}
Let $M =\Upsilon_{{\bf x},1-x_3}$,
in which case the constraints (\ref{nasc1}) are equivalent to
\begin{align} \label{nasc2}
x_1=x_2,\;\;x_i^2 \leq x_3 \qquad (i=1,2).
\end{align}
Then $\frac{1}{2}[I+\sigma_3]$
is a stationary point so that
the channel has an output of rank one.
So the channel:
\begin{align} \label{ch2}
\Upsilon _{{\bf y},u}(\rho)
=(\Delta_\lambda^{(2)}\circ M)(\rho)
=\frac{1}{2}\left[I+ \lambda x_1 w_1 \sigma_1
+ \lambda x_2 w_2 \sigma_2 + \lambda(1-x_3 + x_3w_3)\right]
\end{align}
satisfies the additivity of the MOE (\ref{add}) and 
the multiplicativity (\ref{mult}) for $p\in[1,\infty]$
with $\Omega$ arbitrary.
Note that (\ref{nasc2}) implies that $y_1=y_2$ and
\begin{align} \label{nasc2.1}
y_i^2 \leq y_3(y_3 + u) \qquad (i=1,2),
\end{align}
and that, conversely,
any channel of the form (\ref{ch1}) satisfying
(\ref{nasc2.1}) and $y_1=y_2$
can be written in the form (\ref{ch2}).
In this case (\ref{nasc2.1}) is slightly weaker than
requiring that the shift $(0,0,u)$ is
in the direction of the longest axes of the ellipsoid.
\\
{\bf Example e).}
In this example
we will apply the remark following the lemma rather than
the lemma itself.
Let $\Psi_{\lambda_k}=\Upsilon_{{\bf z},0}$
be the unital qubit channel with
$z_k = \lambda_k$, and
let $M$ be as in the previous example.
With the additional assumption:
\begin{align}
|\lambda_i| \leq |\lambda_3| \qquad (i=1,2)
\end{align}
the optimal inputs of $\Psi_{\lambda_k}$ are
$\frac{1}{2}[I \pm \sigma_3]$.
Therefore, $\Psi_{\lambda_k}$ and $M$
satisfy the conditions in the remark
so that the channel:
\begin{align} \label{ch3}
\Upsilon_{{\bf y},u}(\rho)
&=(\Psi_{\lambda_k}\circ M)(\rho) \notag\\
&=\frac{1}{2}\left[I+ \lambda_1 x_1 w_1 \sigma_1
+ \lambda_2 x_2 w_2 \sigma_2 + \lambda_3(1-x_3+x_3w_3)\right]
\end{align}
satisfies the additivity of the MOE (\ref{add}) and 
the multiplicativity (\ref{mult}) for $p\in[1,\infty]$
with $\Omega$ arbitrary.
Note that $\Upsilon_{{\bf y},u}$ has
the form (\ref{ch1}) with parameters
which satisfy (\ref{nasc2.1}) and
\begin{align} \label{additionalc}
(y_1-y_2)^2 \leq \frac{y_3}{y_3+u}(1-y_3-u)^2.
\end{align}
Conversely,
any channel of the form (\ref{ch1}) which satisfies the condition
(\ref{nasc2.1}) and (\ref{additionalc}),
which are stronger than (\ref{nasc1}),
can be written in the form (\ref{ch3})
with parameters of the channel
$M=\Upsilon_{{\bf x},1-x_3}$
chosen so that $x_1^2=x_2^2=x_3$.
This example is of most interest when
$y_1 \neq y_2$ since
it cannot be written in the form (\ref{ch2}).
A channel with $y_1=y_2$ satisfying
(\ref{nasc2.1}) and (\ref{additionalc}) with strict inequality
can be written in the form (\ref{ch2}) with
$x_i^2<x_3$ or
in the form (\ref{ch3}) with 
$x_i^2=x_3$ and $0<\lambda_i<\lambda_3$
for $i=1,2$.

Finally let us remark that
one can compose these $M$ with
other unitarily covariant channels $\Psi$
such as the transpose depolarising channel to get channels:
\begin{align}
\Phi(\rho)=
(\Psi\circ M)(\rho)=\lambda M(\rho)^{\rm T} +(1-\lambda)\frac{1}{d}I.
\end{align}
Here $\lambda\in [-1/(d-1),1/(d+1)]$,
${\rm T}$ is the transpose.
Note that the additivity of the MOE (\ref{add}) 
for the transpose depolarising channel
was proven [8],[9].
In addition,
composition with the Werner-Holevo channel 
yields another class of channels of the form
$\big(I-  [M(\rho)]^T \big) /(d-1)$.
The additivity of the MOE (\ref{add}) 
and the multiplicativity (\ref{mult})
of these channels were proven in [11]
by a different method.
The channel $M$ given by (\ref{shift}) in this example
was called "stretching" there.

We have shown how to extend
the known results about the additivity of the MOE
and the multiplicativity of the maximal output $p$-norm
to a much larger class of channels,
including many non-unital and asymmetric ones.

{\bf Acknowledgement.}
I would like to thank my supervisor Yuri Suhov
for suggesting the problem,
constant encouragement and numerous discussions.
Mary-Beth Ruskai is thanked for
reading the draft of this note carefully and
giving us important remarks and many examples.
Christopher King and Nilanjana Datta helped us with
some useful discussions.

\end{document}